# Electrospun nanofibers in cancer research: from engineering of *in vitro* 3D cancer models to therapy

Marta Cavo,[†,1] Francesca Serio,[†,1,2] Narendra R. Kale,[3] Eliana D'Amone,[1] Giuseppe Gigli,[2,1] Loretta L. del Mercato*,[1]

[1]Institute of Nanotechnology, National Research Council (CNR-NANOTEC), c/o Campus Ecotekne, via Monteroni, 73100, Lecce, Italy.

[2]Department of Mathematics and Physics "Ennio De Giorgi", University of Salento, via Arnesano, 73100, Lecce, Italy.

[3]MIT-WPU, School of Pharmacy, Kothrud, Pune 411038, India.

† These authors contributed equally.
*corresponding author: loretta.delmercato@nanotec.cnr.it

**Abstract**

Electrospinning is historically related to tissue engineering due to its ability to produce nano-/microscale fibrous materials with mechanical and functional properties that are extremely similar to the extracellular matrix of living tissues. General interest in electrospun fibrous matrices has recently expanded to cancer research both as scaffolds for *in vitro* cancer modelling and as patches for *in vivo* therapeutics delivery. In this review, we examine electrospinning by providing a brief description of the process and overview of most materials used in this process, discussing the effect of changing the process parameters on fibers conformation and assemblies. Then, we describe two different applications of electrospinning in service of cancer research: firstly, as three-dimensional (3D) fibrous materials for generating *in vitro* pre-clinical cancer models; secondly, as patches encapsulating anticancer agents for *in vivo* delivery.

## 1. Introduction

Electrospinning (ES) is a worldwide recognized process to produce fibrous and porous two-dimensional (2D) and three-dimensional (3D) materials starting from a polymeric solution.[1–4] Success of ES is mainly related to its versatility, despite being a simple and cost-effective technique: during the ES process, an electrified polymeric jet experiences bending instability and solidifies to produce long and continuous nanofibers with diameter ranging from a few tens of nanometers to a few micrometers.[5]

ES has been significantly used for tissue engineering (TE) applications, since a huge number of different structures, morphologies and compositions can be obtained and adopted to make fibers suitable for vascular, bone, neural, tendon and ligaments regeneration.[6–14] In particular, a lot of work have been published in the field of bone regeneration, where the nanofibrous substratum provides favourable conditions for bone-associated cell anchorage and growth;[15,16] the random nature of fiber deposition produces surface roughness suitable for cell attachment in bone scaffolds, as well as for osteoblastic differentiation and mineralization that are better regulated on nanofibrous surfaces than on more dense structures.[17] ECM-like electrospun structures are also suitable for muscle cell adhesion, since myoblast cells require highly aligned (anisotropic) or unidirectionally oriented fibers to adhere and proliferate.[18,19] The use of electrospun scaffolds as substrates for neural tissue engineering was also investigated; in this context, several groups have demonstrated how electrospun fibers had great efficacy in enhancing nerve regeneration.[20,21] Last but not least, electrospinning has been used to produce thin tubes to be used as scaffolds to make blood vessels.[22–26]

In cancer treatment, nanofibers have many advantages such as, ability to obtain fibers with diameters ranging from nanometer to sub-micrometer, surface modification, alignment variation and drug encapsulation. Anticancer drug loaded ES nanofiber provide controlled and sustained drug release at desired site of action with improved efficacy. ES nanofibers were also used as an implant into the post-operative tumour cavity, to inhibit tumour recurrence and to prolong drug release at tumour site.[27] Further, ES nanofibers provides unique opportunity to build cell environments which can mimic the *in vivo* tumour micro-environment.[28] Some of the recent work towards 3D *in vitro* cancer model development is mentioned below.

During last years, the research community has focused on the possibility to translate tissue engineering principles and methods into cancer research, with the aim to provide physiologically relevant 3D *in vitro* cancer models. This new branch is known as *Cancer tissue engineering* and aims to explore cancer pathogenesis and evolution in a more realistic environment than 2D cultures or animal models.[29–31] In this context, also the electrospinning technique has been repurposed to obtain 3D fibrous materials with properties similar to those found in native tumours.[32] At the same time, in the last years electrospun matrices have gained attention also in clinical applications: thanks to their porous structure and ability to incorporate drugs within the fibers lumen, electrospun fibers matrices have been used as transdermal drug delivery system (TDDS) to facilitate the delivery through the skin in a controlled way; this application is particularly relevant in case of anticancer treatments that are often associated with several drawbacks, such as poor solubility and instability in the biological environment, adverse effects on healthy tissues and low concentration at tumour site.[33] We have listed some of polymers used for anticancer drug delivery *in vitro* analysis using cancer cell lines (Table 1).

**Table 1** – List of polymers used for *in vitro* anticancer drug delivery studies.

| Polymer | Fiber diameter | Drug | Cancer type | Ref. |
|---|---|---|---|---|
| PLGA | 430 nm | Doxorubicin hydrochloride, and Camptothecin | HepG-2 cells (Human liver cancer) | [34] |
| PLLA | 300 nm | 5-Fluorouracil and Oxaliplatin | Colorectal cancer | [35] |
| PEG-PLLA | 690 nm | Carmustine | Glioma C6 cells | [36] |
| PCL | 80–120 nm | 5-Fluorouracil and Paclitaxel | TNBC cells (Human triple negative breast cancer) | [37] |
| pNIPAM | 600-800 nm | Doxorubincin | Hela cells (Human cervical cancer) | [38] |
| PVA | 260 nm | Doxorubincin | SKOV3 cells (Ovary cancer) | [39] |

*Abbreviations: PLGA-poly lactic-co-glycolic acid, PLLA- poly(L-lactic acid), PEG-PLLA-poly(ethylene glycol) – poly(L-lactic acid), PCL-poly(ε-caprolactone), pNIPAM-poly(N-isopropylacrylamide), PVA- poly(vinyl alcohol).*

However, there are few challenges in ES nanofibers applications such as uniform drug distribution in ES fibers, drug compatibility in polymer solution, to avoid initial burst release of drug, to develop 3D scaffolds with desire porosity. Due to these critical parameters, all the electrospun nanofiber-based systems for cancer research are still in preclinical trials only. We have discussed ahead challenging parameters of ES fibers in cancer research.

All these facts highlight the multiple applications of electrospinning at the service of a problem that is as present as cancer research. The first part of this review will be dedicated to summarize the state of the art of electrospinning-based 3D *in vitro* tumour models, subdivided according to the type of cancer studied or as membranes for intra- and extravasation studies; the second part will review the clinical applications of electrospun fibers for local therapy, giving overall a complete picture of electrospinning meeting cancer research, both *in vitro* and *in vivo*.

## 2. Overview of electrospinning technique

Electrospinning is basically an electro-hydrodynamic phenomenon: it begins with exposure of charged polymer to electric field between the metal needle and the collector; this causes instability within the polymer solution that deforms the spherical droplet to a conical shape (Taylor cone); at this stage, ultrafine fibers are extruded from the conical polymer droplet, which are collected on a metallic collector kept at an optimized distance.[6,40,41] Fiber formation process begins with jet initiation and followed by jet whipping instability. A stable charged jet initiates when applied voltage exceeds a critical value and the electrostatic force overcome fluid surface tension. Charge forces cause the whipping of the liquid jet towards the collector to stretch up to few nanometer diameter. During this process, jet undergoes non-axisymmetric (with reference to jet center line) instabilities, i.e. bending instability and axisymmetric instability i.e. Rayleigh instability. Rayleigh instability take place in presence of two opposing forces on the surface area, i.e. surface tension and electrostatic repulsion of charges in the jet.[5,42] To successfully complete this process, ES apparatus has to be composed of a high voltage source (1–30 kV), a syringe pump to extrude the solution at a defined flow rate, a metallic needle and a conductive collector.[41,43,44] The electrospinning process begins with sufficient high electrostatic force generated by increased applied voltage. Applied voltage is inversely proportional to fiber diameter because of enhanced fiber stretching due to higher voltage. In addition to applied voltage, viscosity of solution has crucial role in ES process. If low viscosity fluid is used, the solution cannot resist to Rayleigh instability and break up into droplets. If viscosity of fluid is sufficient high with long chain molecules content, the solution can withstand against Rayleigh instability and fluid jet leads to nanofiber deposited on the collector.[45,46]

Solvent evaporation plays major role while fluid jet flies towards collector under the influence of electrostatic force. Solvent evaporation from fluid jet occurs prior to get a dry nanofiber deposited on the collector. There should be optimum collecting distance, flow rate and viscosity of solution which is sufficient for solvent evaporation and fiber stretching. Decreased collecting distance and increased flow rate and viscosity result into non-evaporation of solvent, which leads to beads like structures or increased nanofiber diameter. Interplay between applied voltage, solution viscosity, collecting distance, solvent evaporation and flow rate take place during ES process and to get desired nanofiber characteristics, it is necessary to optimize those parameters. Solvent selection is another critical parameter during ES process. Highly volatile solvents may cause jet drying at the tip of the needle because of low boiling point. If polymer dissolves in two solvents, one as solvent and another as non-solvent, ES leads to highly porous nanofibers.[47,48,49]

Although it appears at first as an easy method, the entire process depends on a multitude of parameters, such as: environmental parameters (temperature and humidity), solution properties (viscosity, conductivity and surface tension) and governing variables (distance between the tip and the collector, electrical potential, flow rate, selected polymers, geometry of the collector). All these parameters can affect the capacity of the spinneret, or the fiber and pore diameters or the structure and morphology of the fibers. Notably, through the optimization of the cited parameters, fibers can be obtained randomly or in an ordered way, larger or thinner, with or without beads.

### 2.1. Configurations for electrospinning

The two main principles implemented to achieve ES constructions are as follows,

1. Manipulating the electric field via collector design (auxiliary electrodes):

Electrode arrangements and collector design had been varied to get better alignment of nanofibers. Parallel electrode, multiple electrode and charged pair electrodes had been tested to get desired alignment of nanofibers. Parallel electrode is one of the simplest ways to manipulate electric field and drives the ES jet to swing back and forth between them, which leads to aligned fibers between the electrodes. Fiber should withhold its weight across the gap along with external electrostatic forces. Charged pair electrode have additional forces through alternate application of opposing charges. Alternate switching between charged electrodes causes ES jet to bridge the gap between collector electrodes. Further to improve fiber alignment, steering electrodes used simultaneously along with parallel electrodes collector.$^{50,51}$

2. Moving the collector:

At the beginning, horizontal and vertical ES setups with flat collectors were proposed (Figure 1, a-b). Even though gravitational force affecting the polymer is negligible with respect to the electric field forces, gravity has an effect on the shape of the polymer droplet and the Taylor cone; in particular, the shape of the droplet forming depends on the flatness of the needle tip, so in vertical configuration the needle tip has to be straightened.$^{52}$ Besides that, the real revolution in ES occurred when it was discovered that the geometry of the collector could influence the arrangement of the fibers and, in particular, could allow obtaining aligned fibers instead of random ones. This result can be achieved by substituting the traditional flat collector with a rotating collector (Figure 1c).

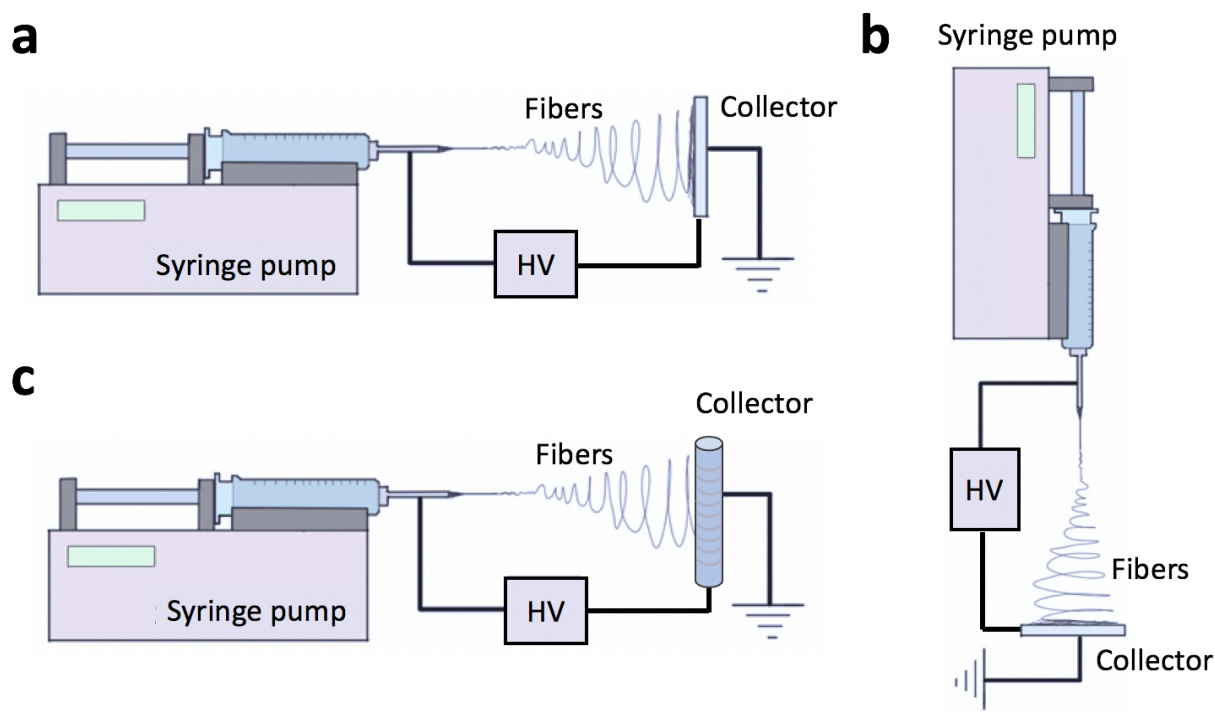

*Figure 1 – Different configurations of ES set-up: horizontal (a), vertical (b) and with rotating collector (c).*

The purpose of the rotating collector is to mechanically stretch the fibers helping them to align along the collector. Collector configurations include a solid cylinder which can rotate about its axis, a conducting wire wound on an insulated cylinder, a wired drum or a disc collector (Figure 2, a-d). With this last configuration, double- and triple-layer highly aligned crossbar structures were obtained.[53]

Also in case of rotating collectors, the alignment of fibers can be optimized by varying several parameters like polymer viscosity and rotational speed, or improving the set-up with two oppositely placed needles as shown by Pan et al..[54] In this work, fibers coming out of the two needles combined in a yarn, which was wound by a cylinder collector rotating at a high speed. Fibers manufactured by this method were continuous, well-aligned, and could be deposited over a large area; moreover, increasing of velocity improved the alignment of the fibers (Figure 2, e-f).

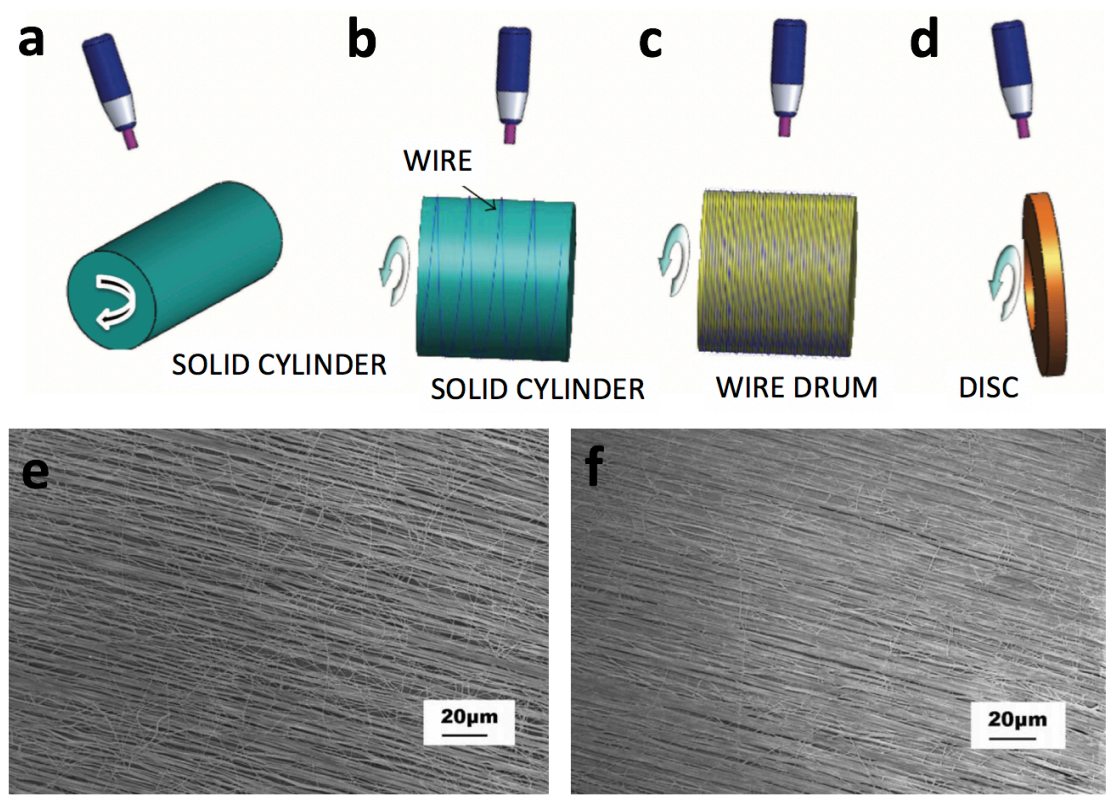

*Figure 2 – Category of ES set-ups based on rotating device. These configurations include (a) solid cylindrical, (b) wire winded on an insulated cylinder, (c) wired drum, (d) disc collector.[53] Well-aligned PVA fibers collected on a Teflon tube with different surface velocity: (e) about 1.3 m/s, (f) 2.3 m/s.[54] Reproduced from ref. 54 with permission from [Elsevier], copyright [2006].*

## 2.2. Materials

So far, more than fifty different polymers and solvents have been successfully electrospun into fibers with diameters in the range from 3 nm to 1 mm. The interactions between solvent and polymer have been found to be crucial in ensuring a successful production of continuous nanofibers.[55] In addition, the solvent selection is important for solution conductivity and may affect the fiber size distribution.[56] Substantially, the material choice depends on the application of the scaffold. For TE purposes, the choice is based on the properties of the tissue to be regenerated and biodegradable polymers are preferably used. The mostly synthetic materials used in TE are polyesters, such as poly(lactide) (PLA), poly(glycolyde) (PGA) and poly(caprolactone) (PCL). Natural materials, such as cellulose, collagen, natural silk, fibrinogen, chitosan and hyaluronic acid are of interest for use in regenerative medicine because the material becomes more recognizable for the cells.[57] However, not all these natural polymers can be easily used with ES: for example, processing chitosan with ES is difficult for its polycationic nature and its high viscosity in solution, which needs toxic or highly acid solvent to electrospin;[58] again, gelatin is not stable if electrospun as single material; finally, hyaluronic acid solutions have very high viscosity and consequently producing uniform fibers from them can be difficult.[59]

The available literature in this field demonstrates the ability to electrospin natural polymers with the possibility to exploit their biocompatibility.[60] However, when compared with the synthetic materials, two main drawbacks such as mechanical performance and difficult electrospinnability are clear.

### 3. Electrospinning for 3D *in vitro* cancer models

Traditionally, cancer biology research has involved *in vitro* analysis of cell behaviour predominately using two-dimensional (2D) cell cultures and *in vivo* animal models[61]: in detail, 2D models are routinely used as initial systems for evaluating the effectiveness of molecules as potential therapeutic drugs; this initial screening precedes animal studies before advancing to human clinical trials.[61] The dissimilarities in cell behaviour between 2D cultures and real tumours are well known and they mainly derive from changes in gene expression originated from the different interactions to which cells are subjected within a 2D microenvironment if compared to a more natural 3D.[62–70] A striking example of that is represented by the unequal nutrient concentration to which cells are exposed: in 2D cultures cells are uniformly exposed to nutrients, while *in vivo* the concentration of soluble factors influencing cell proliferation is characterized by spatial gradients that play a vital role in biological differentiation, organ development and countless other biological processes.[71–74] Cancer cells would alter their behaviour in 2D culture from their *in vivo* behaviour. However, 3D scaffolds will mimic *in vivo* conditions which will help to precise analysis of cellular response against anticancer drugs. It is therefore not surprising that many aspects of tumorigenesis are still not fully understood.[75] Below, we have listed noted difference of cell line behaviour on 3D scaffolds than on 2D culture of same material (Table 2).

Table 2 – List of nanofiber scaffolds used for 3D culture technique with noted difference of cell line behavior on 3D scaffolds than on 2D culture of same material.

| 3D nanofibers scaffolds | Fiber diameter | Cell line | Noted difference from 2D culture | Ref. |
|---|---|---|---|---|
| PCL | 400 nm to 2 μm | CT26 colon cancer cells and bone marrow-derived dendritic cells (BM-DC) | Lipopolysaccharide (LPS)-activated BM-DCs showed increased expression of CD86 and major histocompatibility complex Class II | 76 |
| PEG | < 600 nm | NIH 3T3 cells | Five times higher cell proliferation | 77 |
| PCL | 295-701 nm | Triple negative breast cancer (TNBC) cells | Higher mammosphere forming capacity and aldehyde dehydrogenase activity, higher cell proliferation and elongation | 78 |
| | --- | Ewing sarcoma cells | More resistant to traditional cytotoxic drugs, also exhibited remarkable differences in the expression pattern of the insulin-like growth factor-1 receptor/mammalian target of rapamycin pathway | 79 |
| silk | 497 nm | HN12 cells | Increased Paclitaxel drug concentration in orders to get cytotoxic effect | 80 |
| PVA and PEOT | 5–10 μm | Pancreatic ductal adenocarcinoma cells (PDAC) | Primary PDAC cells showed good viability and synthesized tumour-specific metalloproteinases (MMPs). such as MMP-2, and MMP-9 | 81 |
| PDMS and PMMA | 1.15-3.6 um | Human lung cancer epithelial cells (A549) | Leads to the formation of cell spheroids from single cells | 82 |

*Abbreviations: PCL - poly(ε-caprolactone), PEG - poly(ethylene glycol), PCL - poly(ε-caprolactone), PVA - poly(vinyl alcohol), PEOT - poly(ethylene oxide terephthalate), PDMS - Poly(dimethyl siloxane), PMMA - poly(methylmethacrylate).*

On the other side, there is a growing awareness of the limitations of animal research and its inability to make reliable predictions for human clinical trials. Indeed, animal studies seem to overestimate by about 30% the likelihood that a treatment will be effective because negative results are often unpublished.[83] For all these reasons there is a growing interest in developing new 3D *in vitro* cancer models to be adopted as pre-screening models and able to recapitulate the 3D microenvironment where cancer cells live.

### 3.1. Electrospinning as extracellular matrix (ECM) of cancer microenvironment

Cancer cells within a solid tumour are in close contact with ECM, that is mainly composed of collagen, fibronectin, elastin, laminin and proteoglycan[84,85] with additional components for specific tissues.[86] In a pathological environment like the cancer one, ECM cannot be considered just as a passive environment surrounding cells, but as an active component with a critical role in regulating tumour cell properties and behaviour.[87] Specifically, in breast cancer ECM stiffness affects cancer development and migration;[73,88,89] in ovarian cancer patients, the expression of collagen VI correlates with tumour grade and it has been shown to be upregulated in drug-resistant cancer cells, which can subsequently remodel the ECM to further promote chemoresistance;[90] again, in small cell lung cancer (SCLC) the adhesion of cancer cells to ECM enhances tumorigenicity and confers resistance to chemotherapeutic agents.[91] Ideally, a 3D tumour model should be able to recapitulate this cell-ECM crosstalk. In order to study the ECM-cancer cell interactions in a 3D environment, biomimetic substrates have been produced.[92] A number of strategies exist, above all the use of Matrigel, a commercial basement membrane preparation.[93–95] Besides that, a lot of 3D scaffolds have been used to culture cancer cells *in vitro*, allowing for their growth and proliferation in a context closer to reality than 2D traditional cultures. Electrospun fiber scaffolds provide an effective artificial 3D culture matrix due to their biomimicry of the architecture of ECM[96,97] and high versatility in biochemical stimuli, including growth factors, adhesion molecules, and drugs.[60,98] In the following paragraphs we will analyse the state of the art of *in vitro* 3D tumour models with a focus on electrospun micro- and nanofibrous, citing some of the most relevant works in the field. The paragraphs have been subdivided according to the cancer type.

### 3.1.1. Breast cancer

Breast cancer is the most common cancer in women across most ethnic groups and accounts for 30% of all new cancer diagnoses.[99] Consequently, a lot of interest in treating it exists. Among all the 3D *in vitro* breast cancer models proposed in the literature, some based on ES technique showed very promising results. In 2012, Saha et al. created electrospun fibrous scaffolds of PCL dissolved in 1,1,1,3,3,3-hexafluro-2-propanol (HFIP) with random and aligned fiber orientations in order to mimic the structure of ECM; human breast cancer cell line MCF-7 was cultured on fibrous scaffolds for 3-5 days, showing elongated morphology in the aligned fibers and maintaining a mostly flat stellar shape in the random fibers, demonstrating how the topographical cue may play a significant role in tumour progression.[100] In a different work, Girard and colleagues showed that 3D scaffolds prepared by ES promoted cancer cell growth in irregular aggregates similar to *in vivo* tumoroids with epithelial to mesenchymal transition (EMT) induction, as shown by upregulation of vimentin and loss of E-cadherin expression.[101] In another work by Du et al., the expression of the adhesion-related genes in human breast cancer cells cultured in electrospun nanofibrous scaffolds was time-dependent.[84,102] In 2015, Guiro et al. fabricated scaffolds made of poly (ε-caprolactone) (PCL) having aligned or random fibers. Proliferation, viability and cell cycle analyses of breast cancer cells seeded on these substrates indicated that this 3D culture system prompted the more aggressive cells to adopt a dormant phenotype. The findings indicate that random and aligned fibrous PCL scaffolds may provide a useful system to study how the 3D microenvironment affects the behaviour of breast cancer cells.[103] In 2017, Rabionet et al. fabricated through ES scaffolds from different PCL-acetone solutions. PCL meshes were seeded with triple negative breast cancer (TNBC) cells. Notably, under the tested conditions, cells exhibited higher mammosphere forming capacity and aldehyde dehydrogenase activity than 2D cultured cells.[78]

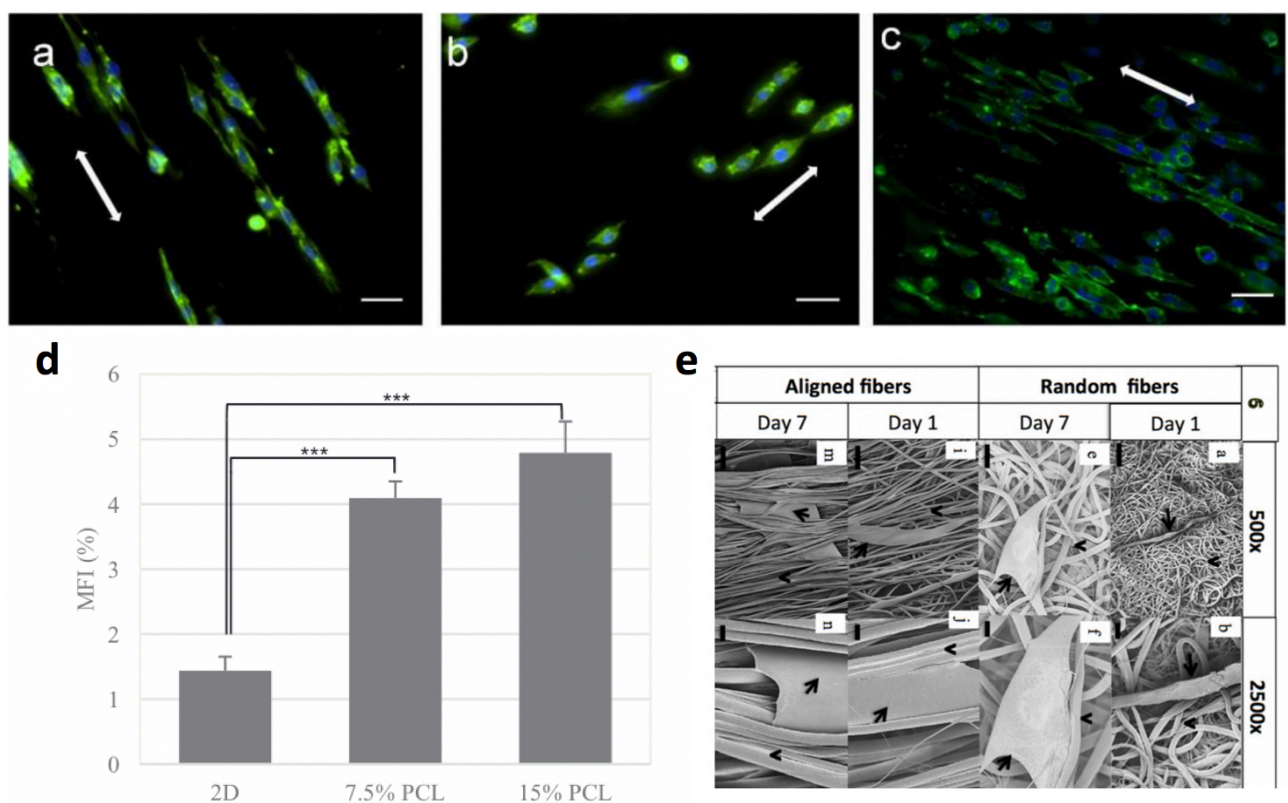

*Figure 3 – Response of different breast cancer cell lines on aligned PCL fibers: (a) 4T1, (b) MTCL, (c) MDA-MB-231. Cells were cultured on PCL fibers for 3 days. Green represents actin cytoskeleton and blue the nuclei of a cell, scale bar = 40 μm.[98] (d) Mammosphere Forming Index (MFI) of MCF-7 cells after 2D and 3D culture on electrospun scaffolds. Statistical significant level was p<0.001.[102] (e) SEM images of MDA-MB-231 cells on fibrous scaffolds after day 1 and day 7 of culture. The arrows depict the cell body and the arrowheads depict the fibers.[101][101] Reproduced from ref. 98 with permission from [John Wiley and Sons], copyright [2017], and ref. 102 with permission from [Elsevier], copyright [2011].*

### 3.1.2. Pancreatic cancer

Pancreatic ductal adenocarcinoma (PDAC) is a highly lethal neoplasm, with a 5-year survival rate of only 6%.[104] These mortality rates depend on several factors: lack of diagnostic markers, high degree of infiltration and metastasis and strong resistance to conventional chemotherapy. Moreover, PDAC is characterized by a very heterogeneous and rich stroma, comprising cellular and extracellular matrix (ECM) components, which seems to play a crucial role in the inefficiency of chemotherapy.[105] For all these reasons, there is a strong need of new 3D *in vitro* PDAC models, even though few electrospun-based were proposed. Among these, in 2013 He et al. successfully engineered a subcutaneous model using an electrospun scaffold[106] and, few years later, the same group proposed a metastatic orthotopic pancreatic tumour model that recapitulated the tumour formation and a hepatic metastasis (the landmark event of pancreatic cancer) by using a polyglyconate/gelatin electrospun scaffold. This metastatic tumour model showed an increased incidence of tumour formation, an accelerated tumorigenesis and a significant hepatic metastasis.[107] In 2014, Ricci and colleagues analysed the interactions of primary PDAC cells with different polymeric scaffolds, respectively characterized by sponge-like pores versus nanofiber mesh interspaces, demonstrating that PDAC cells show diverse behaviour when interacting with different scaffold types that can be exploited to model various phases of pancreatic tumour development and invasion.[81]

### 3.1.3. Colorectal cancer

Electrospun fiber mats have been also exploited for providing preclinical 3D *in vitro* models of colorectal cancer, the third most common cancer among both men and women worldwide. In 2013, Yamaguchi et al., attempted to grow 3D cultures of various tumour cell lines, including human colon adenocarcinoma HT-29 and DLD-1 cells, on silicate fibers (SNFs) prepared by ES technology via a sol-gel process.[108] HT-29 and DLD-1 cells seeded on SNFs scaffolds grew gradually and formed tight aggregates showing 3D cell morphology. Notably, cell viability assay for HT-29 cells showed a significant increased drug resistance to mitomycin C in 3D cultures, likely because the drug penetration was restricted in the interior of the 3D structure thus reducing the drug exposure in cells. Overall, the study demonstrated that SNF fibers obtained by ES technology could be used as *in vitro* 3D tumour model for testing the efficacy of potential anticancer drugs. In 2016, Kim et al. explored electrospun mats made out of intertwined PCL nano- and submicron-scale fibers to grow 3D cocultures of colon cancer cells and dendritic cells. The authors reported the cocultures of mouse CT26 colon cancer cells and bone marrow- derived dendritic cells (BM-DCs) on PCL fibers characterized by efficient adhesion, spreading, and migration of top-seeded cancer cells and BM-DCs on the surface of the mats. In particular, they showed that PCL mats composed of a mixture of nanofibers with diameters ranging from 400 nm to 10 μm permitted an optimal cell infiltration. Notably, BM-DCs cocultured on 3D PCL mats maintained their ability to sprout cytoplasm, to migrate, to synapse with and to engulf mitoxantrone-treated CT26 cancer cells thus reproducing the *in vivo* cross-talk between these two of cancer and immune cells.[76]

### 3.1.4. Prostate cancer

In 2009, Hartman et al. created electrospun micro- and nanofibrous collagen scaffolds to support prostate tumour growth in 3D. The effects of matrix porosity, fiber diameter, elasticity and surface roughness on growth of cancer cells were evaluated. Obtained data indicated that while cells attach and grow well on both nano- and microfibrous electrospun membranes, the microfibrous membrane represented a better approximation of the tumour microenvironment. It was also observed that C4-2B non-adherent cells migrated through the depth of two electrospun membranes and formed colonies resembling tumours on day 3. An apoptosis study revealed that cells on electrospun substrates were more resistant to both anti-neoplastic agents, docetaxel (DOC) and camptothecin (CAM), compared to the cells grown on standard collagen-coated tissue culture polystyrene (TCP).[32]

### 3.1.5. Ewing sarcoma

Ewing sarcoma is a bone tumour mostly prevalent in adolescents and young adults that is rapidly fatal unless effectively treated.[109] Considering most preclinical studies fail to predict whether a given drug candidate will work in clinical trials, in 2013 Fong et al. developed an ex vivo model based on PCL 3D mats, which conferred a more physiologically relevant cell phenotype compared with traditional 2D cultures. Moreover, this model demonstrated pronounced upregulation of the insulin-like growth factor-1 receptor (IGF-1R) pathway, a receptor commonly expressed in Ewing sarcomas and one of the most promising targets in drug development phases.[79] Two years later, the same group improved the model to better mimic *in vivo* biology and drug sensitivity; in this work, Ewing sarcoma cells were seeded onto electrospun scaffolds within a flow perfusion bioreactor, where the fluidic shear stress could provide a physiologically relevant mechanical stimulation. They found that cells exposed to a fluidic flow produced more IGF1 ligand, demonstrating shear stress-dependent sensitivity to IGF-1 receptor targeted drugs as compared with static conditions. Moreover, flow perfusion increased nutrient supply, resulting in an enrichment of cell culture over static conditions (Figure 4).[110]

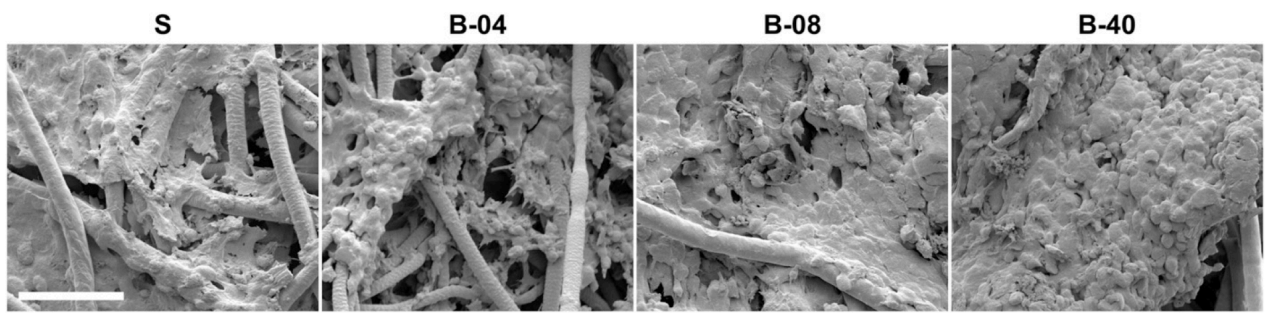

*Figure 4 - SEM micrographs of Ewing sarcoma cell culture under static and flow perfusion conditions (S, static; B-04, 0.04 mL/min; B-08, 0.08 mL/min; and B-40, 0.40 mL/min). After 10 days of culture, scaffold surfaces for the different experimental groups were examined by SEM (scale bar, 50 μm).*[110]

### 3.1.6. Glioblastoma

Glioblastoma multiforme is the most aggressive and malignant brain tumour.[111] Available treatments are ineffective and some tumours remain inoperable because of their size or location. With the aim to advance the research in this field, in 2014 Jain et al. investigated the apoptosis of glioblastoma tumour induced by directional migration of tumour cells through the aligned fibers of an electrospun scaffold to a gel pool. It is difficult to treat brain cancer with current modalities due to drug delivery challenges at the desired site of action. However, if one could control tumour cell migration, it would be useful approach in cancer therapeutics. Jain et al. demonstrated enhanced cell migration towards cyclopamine conjugated hydrogel sink, by using aligned nanofiber conduits. Results showed that this conduit with aligned fibers could promote the migration of glioma cells from the glioma tumour, finally providing an innovative method which may open up a new way for the application of electrospun fibers in the treatment of cancer.[112]

### 3.1.7. Lung cancer

In 2013, Girard et al. developed 3D nanofibrous scaffolds by ES a mixture of poly(lactic-co-glycolic acid) (PLGA) and a block copolymer of polylactic acid (PLA) and mono-methoxypolyethylene glycol (mPEG). The ability of cancer cells from different tumours to form aggregates on these scaffolds was investigated. Among these, LLC1 lung cancer cells were seeded on scaffolds and monitored for morphology and cell viability. LLC1 cells when cultured on a 3P scaffold formed tumoroids at day three. Cells showed different characteristics in 3D scaffolds as compare to monolayer culture. In 3D culture cells were more resistant to anticancer drug along with upregulation of vimentin and loss of E-cadherin expression.[101]

### 3.1.8. Bladder cancer

Considering the proven critical role of ECM in determining cell behaviour and tumour progression, in 2016 Alfano and colleagues performed a deep characterization of the healthy and tumoral human bladder ECM, finding the main features characterizing the tumoral one: (i) a loss of tissue morphology, (ii) the linearization and degree of organization of fibrils of thin diameter and (iii) an increased vascularization. From these findings, authors modelled a 3D synthetic bladder model able to create an environment more similar to the *in vivo* microenvironmental niche conditions to test innovative therapies. To this aim, they produced electrospun PCL-based scaffolds; results showed that binding and invasion of bladder metastatic cell line were observed on synthetic scaffold recapitulating anisotropy of tumoral ECM, but not on scaffold with disorganized texture typical of non-neoplastic lamina propria.[113]

### 3.1.9. Melanoma

The incidence and mortality rates of melanoma, the most malignant skin cancer, have been rapidly growing in recent years,[114] but the effective management of this disease is still a challenge due to lack of suitable culture systems. Trying to overcome this limitation, in 2017 Wang et al. developed a poly(γ-benzyl-L-glutamate)/poly(lactic acid) (PBLG/PLA) electrospun membrane and demonstrated its suitability as a matrix for 3D culture of melanoma cells. In detail, their results show that – compared to other substrates - PBLG/PLA nanofiber membranes could better support cell viability and proliferation and, besides that, promoted the generation of tumoroid-like structures. These findings demonstrated that the proposed model could mimic the ECM of melanoma microenvironment and be a promising tool for 3D cell cultures.[115]

### 3.2. Electrospinning as membrane for intra- and extravasation studies

All the 3D models presented in the previous section are promising in bridging the gap between *in vitro* and *in vivo* testing; however, they fail to mimic some specific and crucial steps of tumour evolution, including cell motility and metastatic cascade. Considering the lethal impact of these mechanisms, there is a strong need to provide not only realistic *in vitro* models of primary tumours, but also reliable new models for intra- and extravasation, the two main phases of metastasis. National Cancer Institute described extravasation as "the movement of cells out of a blood vessel into tissue during inflammation or metastasis (the spread of cancer)" and intravasation as "the movement of a cell or a foreign substance through the wall of a blood or lymph vessel into the vessel itself." Cancer cells invade the stroma and metastasize to blood vessels as singlets or clusters. This stromal invasion is followed by extravasation into and intravasation from the blood vessels and subsequent invasion of neighbouring tissues. Invasive cells then re-establish distal colonies, causing metastasis.[116]

Thanks to their fibril structures, electrospun membranes have been proposed to mimic the tumour-blood vessel interface, the physical zones where intra- and extravasation take place, or specific zones where metastasis can happen, such as bone. Some examples of hybrid systems are already present in the literature. For instance, Cavo and colleagues combined an alginate-Matrigel based gel with embedded breast cancer cells with electrospun membranes to see cell invasion capability; results showed that cells were able to migrate through the gels and attach to the membrane mimicking the vascular walls hosted within a bioreactor.[72] In a different work, Shokoohmand et al. successfully grew human bone-derived osteoblast cells onto electrospun PCL scaffolds to mimic the bone niche and integrated it into a bioengineered 3D *in vitro* model recapitulating prostate cancer metastasis to the bone.[117] Again, Rabionet et al. developed and used PCL scaffolds for 3D breast cancer cell culture by using ES nanofibers.[78]

Despite the little supporting literature, combining 3D tumour models with cellular infiltration throughout nanofiber scaffolds seems to be crucial for realizing complete engineered cancer models and to better recapitulate the microenvironment of cancer cells.

## 4. Electrospun nanofibrous devices for cancer therapy

Cancer treatment represents one of the most crucial issues in clinical management. Uncontrolled growth, immortality and the ability to metastasize are significant characteristics of cancer cells.[118] This paragraph aims to explore the existing applications of electrospun nanofibers in cancer therapy, mostly focusing on drug/gene delivery, cancer cell detection/sensing and highlighting the challenges and future directions for applications of electrospun nanofibers in cancer research.

### 4.1. Drug delivery

Conventional surgery, radiotherapy and chemotherapy are currently the most adapted treatment modalities for cancer.[119] However, these options are often limited and inadequate.[120] Actually, the clinical use of the most engaged chemotherapeutics against several solid and hematopoietic cancers, including breast cancer, osteosarcomas, aggressive lymphomas, and leukaemia, is usually limited by the severe and harmful side effects at therapeutic concentrations.[63] Moreover, in order to maximize the therapeutic effects, many patients need to take excessive amount of drugs, orally or via systematic injection, which might trigger side effects in healthy tissues.[121]

In order to limit the common systemic side effects and toxicity, researchers are currently focusing their work particularly in postoperative chemotherapy applications for developing innovative and sophisticated treatments to specifically target cancer cells.

Targeted microcarrier technologies such as micro- and nanoparticles, liposomes and nanofibrous materials are designed for drug delivery in a controlled manner to a specific site. The release of drugs in a controlled way is a complex challenge in which systems provide an optimal amount of drug to a specific target at a predetermined rate and for a definite time period (ranging from days to years).[122]

Controlled release systems offer advantages if compared to conventional drug therapies. Firstly, they maintain the drug in the desired therapeutic range in the blood by a single administration and avoid oscillating drug levels attributed to the standard trend of injected drugs in the blood (the blood level of the drug rises, reaches a peak plasma level and then declines) which cause alternating periods of ineffectiveness and toxicity. Other advantage comprises the localized drug delivery to a specific target of the body. In this manner, one can guarantee an improvement of the pharmacological properties of free drugs and an increment of the patient compliance with a reduced amount of needed drug and a decreased frequency of drug administration.[122–124] Moreover, contrary to sustained release formulations such as suspensions and emulsions which are influenced by the environmental conditions and therefore are subjected to patient variations,[122] controlled release system is provided if the drug is reserved in a polymeric material. The release of the drug through the polymeric system network is controlled by two main mechanisms: 1) drug diffusion, which is the most common release mechanism whereby the drug moves in a concentration gradient from the inner part to the system's outer side up to the body; and 2) chemical mechanisms, including the degradation of the polymer or the cleavage of the drug from a polymer linkage to allow the drug release.[122]

Electrospun fibers represent an ideal platform for drug delivery since they provide high surface area to volume ratio and may favour abundant drug release.[125] Various delivery systems have been developed by loading antibiotics, anticancer drugs, proteins and also nucleic acid. The application of electrospun fibers for chemotherapy became popular more recently, in particular as local drug delivery system after a surgical operation for removing solid tumours.[121]

The active molecules can be (i) physically or chemically attached to the surface of the fibers or (ii) mixed with the polymers. In the following, the different loading methods will be presented.

### 4.1.1. Drug loading

Drug solubility in the polymer solution is decisive for selecting the right loading method. The drug loading method, for its part, has a pivotal role on release process. Generally, hydrophobic drugs such as doxorubicin and paclitaxel are well processed with organic solvent.[126,127] Hydrophilic drugs, such as peptides and proteins are better handled in water-soluble polymer.[128,129] Here, the several possible routes for drug loading will be discussed.

The ES process with two or more constituents is one major approach to produce nanofibers encapsulating pharmaceuticals or other bioactive components with a significant drug loading ability. This configuration require the application of the conventional ES device (a single nozzle with a single capillary) in which both drugs and polymers are mixed together in the same solvent and then electrospun into drug-loaded fibers.[130–133]

Ranganath et al. reported the successful production of electrospun poly-(D,L-lactide-co-glycolide) fibers to deliver paclitaxel for postsurgical chemotherapy against malignant brain tumours. The device showed sustained paclitaxel release over 80 days *in vitro* and demonstrated tumour growth inhibition in animal study.[134] Xu and co-workers developed implantable biodegradable system for controlled release of 1,3-bis(2-chloroethyl)-1-nitrosourea anticancer drug by ES a copolymer solution. The drug resulted well dispersed through the polymer matrix and its release appeared directly proportional to the loaded drug amount.[36] Liu et al. designed biodegradable drug carriers for in situ treatment of liver cancer based on electrospun PLLA nanofibers doxorubicin hydrochloride-loaded in order to treat locally an unresectable liver cancer or to prevent a post-surgery tumour recurrence avoiding the systemic chemotherapy. The amount of doxorubicin hydrochloride-loaded have been completely released by diffusion from the nanofibers.[135]

Following this approach, Zong et al. developed electrospun cisplatin-loaded delivery system to treat cervix cancer in mice via vaginal implantation. It showed to have good mucoadhesive property and high anti-tumour power thanks to the cisplatin release from the nanofibers.[136]

Ramachandran et al. developed a localized nanoimplant for controlled delivery of anti-cancer drug, Temozolomide, in orthotopic brain-tumour. Specifically, a library of drug loaded (20 wt%) electrospun nanofiber of PLGA-PLA-PCL blends was produced with distinct *in vivo* brain-release kinetics (hours to months). Orthotopic rat glioma implanted wafers showed constant drug release (116.6 μg/day) with negligible leakage into the peripheral blood (< 100 ng) rendering ~ 1000-fold differential drug dosage in tumour versus peripheral blood. Most importantly, implant with one-month release profile resulted in long-term (> 4 month) survival of 85.7% animals whereas 07 day releasing implant showed tumour recurrence in 54.6% animals, rendering a median survival of only 74 days.[137]

However, the ES is not immune from few drawbacks. Above all, this method of synthesis leads to an initial burst release of the drug.[130,138] It is given by fast diffusional effects and probably it is predisposed by the low physical interactions between the drug and the polymer, so that, during the process, the most of the drug molecules are likely to be localized around the fiber surface.[130,139,140]

As a result, most of drug amount resulted in contact with the surrounding water solution and it causes a subsequent large burst release at short times.[139] In addition, the burst release is often affected by the biodegradable feature of chosen carrier.[130–132,140] Thus, a special caution is required to tailor both the release and the degradation rate, for example by varying the polymer blend composition and the ratio of amorphous to crystalline segments in order to control whether drug release occurs via diffusion alone or by diffusion and scaffold degradation.[141,142]

In contrast to ES, researchers have developed original approach to limit the initial burst and, in this sense, to extend the drug release over a long period of time.

A good way to prolong the therapeutic release of drug from electrospun fibers is to add inorganic nanocarrier into the fibers, such as silica based nanoparticles[143–145] and hydroxyapatite.[146]

For example, Zhou et al. implemented the conventional ES setup by firstly loading the anticancer doxorubicin drug into $SiO_2$ nanoparticles and then mixing the resulted solution with the polymer solution.[143] In the study of Zheng et al. the amoxicillin, a model drug, was absorbed to the hydroxyapatite surface particles before to disperse them into PLGA solution ready to be electrospun. The model showed a sustained release profile and a limited initial burst release.[146] Moreover, novel drug carriers with higher affinity for both drug molecules and polymer solution were then engaged. An example is the star polymer ad hoc synthetized in the work of Balakrishnana and co-workers to be optimal for the further applications. The star polymer was blended with the more traditional PCL and the system was used for doxorubicin sustained release.[147]

Another big issue in mixing the drug and polymer together is the exposition of drugs to harsh solvent, often necessary to dissolve the most used polymers. A range of biomolecules including anti-cancer drugs and not only, can be easily immobilized on the surface of electrospun fibers via chemical or physical method for sustained drug delivery. This way of surface functionalization has at least two main advantages if compared to conventional ES. Firstly, physical and chemical drug immobilization after fiber synthesis can avoid the denaturation or inactivation of the drug generally caused by high voltage or harsh organic solvents; secondly, the possibility to add additional materials after the ES process leads to avoid unexpected but usual trouble in term of electrospinnability, for instance the difficulties to dissolve such biomolecules in an organic solvent because of their high molecular weight and the charges present on their surface.

Physical non-specific random adsorption is governed by electrostatic interactions, hydrophobic interactions, hydrogen bonding, and van der Waals interactions between drugs and fiber surface.[148] However, non-specific adsorption allows the formation of weak links and it may favour a significant burst release. This is the reason why it is rarely used for anticancer therapy applications.

More applied is the immobilization after chemical treatment of fiber surface by which is possible to tailor fiber adhesion properties by adding apposite functional groups such as amine, carboxyl, hydroxyl or thiol groups suitable for further immobilization.[149] Chemical conjugation methods are better for finely controlling the amount of incorporated drug on the nanofiber mesh than physically trapping the drug on the surface, and exhibits slow drug release kinetics with reduced initial burst release.[150] For instance, Volpato and co-workers reported a chemical immobilization of heparin-containing polyelectrolyte complex nanoparticles and basic fibroblast growth factor onto chitosan electrospun fibers. They prevented the burst release effect by post-fabrication modification of fiber surface by adding a single bilayer of polyelectrolyte multilayer composed of N,N,N-trimethyl chitosan and heparin.[151] Im et al. limited the initial burst by modifying cross-linked hydrogel fibers through fluorination before the addition of the model drug.[152] A versatile surface modification method is the approach performed by Ma et al. that worked on a porous nanofiber device made of chitosan/polyethylene oxide mixture. After fabrication, nanofibers were soaked into water to remove polyethylene oxide and then immersed in a solution with the anticancer drug to load it. Finally, the porous drug-loaded device was dipped in hyaluronic acid solution with the aim to control drug release thanks to chitosan matrix and hyaluronic acid affinity.[153] The best promising strategy to limit the initial burst release is through core-shell fibers by using coaxial needles.[154–156] Co-axial ES technique has a high potential in biomedical applications for core/sheath nanofibers since the polymer shell generally protects the core compounds from direct exposure to external environment and, consequently, contributes to the prolonged drug release.[157] Coaxial ES is very useful mainly for the encapsulation of bioactive compounds which are susceptible to harsh organic solvents (drugs, proteins, cells and nucleic acids).[155,156] For instance, Zhang and collaborators demonstrated the successful loading of fluorescein isothiocyanate-conjugated bovine serum albumin, a model protein, in a water-soluble core of poly(ethylene glycol) within a PCL shell. These core-shell fibers showed a reduction of the initial burst release compared to that of fibers electrospun from the mixture of the poly(ethylene glycol) and PCL obtained with a conventional setup.[156] Mickova et al. reported the development of nanofiber-liposome system for drug delivery. They tested the incorporation of liposomes containing the horseradish peroxidase drug with co-electrospinning and with coaxial electrospinning and demonstrated that the traditional setup doesn't preserve the integrity of the liposome which, on the contrary, are well reserved. In this second case, the therapeutic activity of encapsulated drug was demonstrated.[158] Another way to slow down the kinetics of release is the development of multilayer stacked nanofibers by ES subsequently two or more different solutions. Generally, these systems combine an initial fast release with a sustained release thanks to layers with different diffusion pathways and degradation behaviours.

Generally, these systems combine an initial fast release with a sustained release thanks to layers with different features. The versatility of layer by layer fiber deposition leads to a range of modification and implementation to make it suitable also for multidrug delivery system. Combining multiple treatments in different time has a key role in chemotherapy for cancer treatment. Okuda and co-workers created a tetra-layered drug-loaded nanofiber meshes for dual release. The construction was made by a first drug-loaded layer, a second barrier mesh, a third drug-loaded layer and a final barrier mesh. An *in vitro* release assay demonstrated that the tetra-layered device can provide timed dual release of the respective drugs by tailoring morphological features of each component mesh like the barrier mesh thickness.[148–150,159]

4.1.2. On-demand drug delivery fibers

Since a number of polymers respond to external stimuli with consequent changes in their physical properties, a new frontier in drug delivery research is the development of smart polymeric fibers which release encapsulated drugs under application of an external stimulus (temperature, light, magnetic field, pH) for an on-off reversible switch.[160–162]

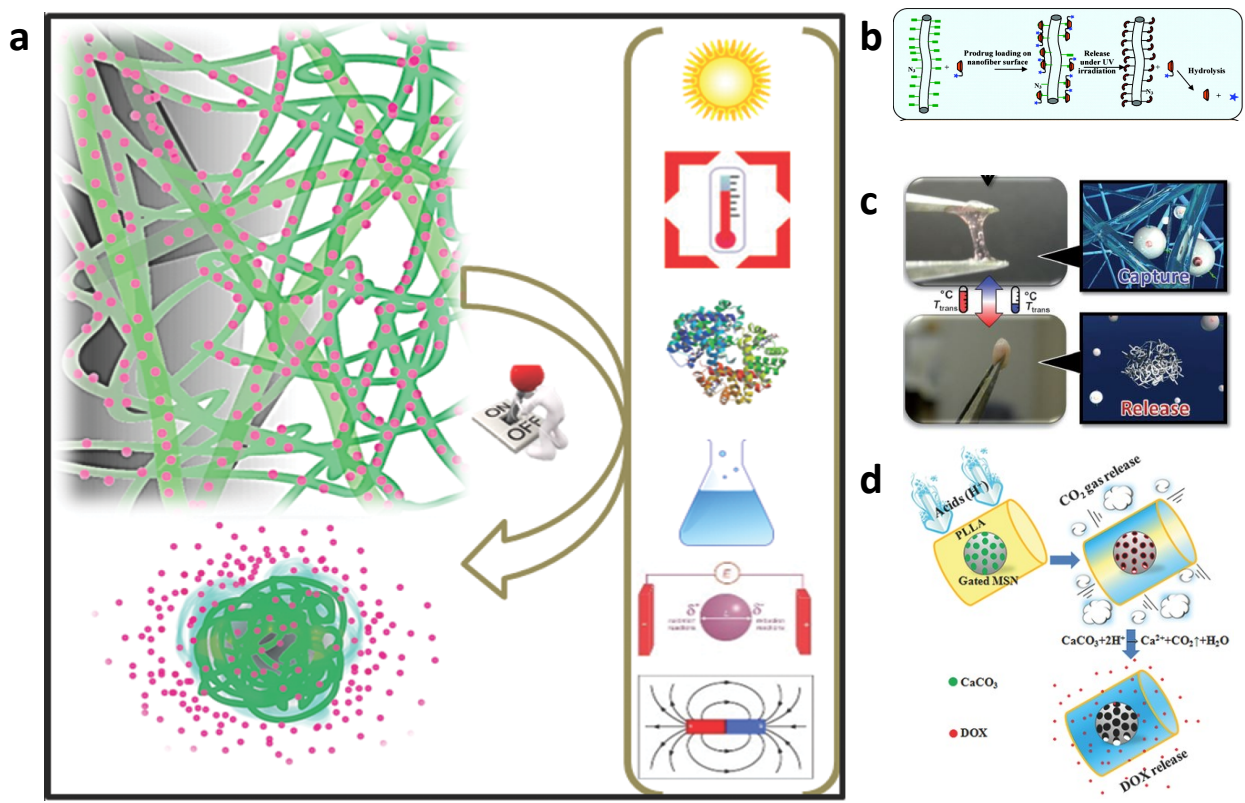

*Figure 5 – (a) Illustration of the mechanisms of on demand drug release under biological, chemical, light, temperature, magnetic, electric field stimuli.[163] (b) Schematic of the photocontrolled drug delivery based on host-guest linkage on the photoresponsive nanofiber surface.[161] (c) A nanofiber*

*web that captures and releases cells by switching to a hydrogel-like structure in response to temperature changes.*[164] *(d) A drug-loaded intelligent electrospun fiber mat which react only in pathological acidic environments by producing $CO_2$ gas with accordingly water penetration into the core of the fibers and rapid drug release.*[165] Reproduced from ref. 161 with permission from [American Chemical Society], copyright [2009], ref. 163 with permission from [John Wiley and Sons], copyright [2014], ref. 164 with permission from [John Wiley and Sons], copyright [2012] and ref. 165 with permission from [John Wiley and Sons], copyright [2015].

In these systems, the drug is localized at a specific targeted area by internal or external forces and then activated. Recently, magnetic particles carrying drug molecules were developed to target the drugs to specific sites in the body by external magnetic fields. Shortly after concentration in the targeted region, the drug molecules were gradually released, thus improving their therapeutic efficiency, while lowering the collateral toxic side effects on healthy cells or tissues.

For instance, Kim and co-workers reported temperature-responsive electrospun nanofibers for 'on–off' switchable release of dextran as drug model.[166] Afterwards, Kim et al. designed the smart nanofibers with bi-functional action by ES a chemically-crosslinkable temperature-responsive polymer with magnetic nanoparticles and anticancer drug (doxorubicin) for induction skin cancer apoptosis. The nanofibers showed simultaneous heat generation and drug release in response to 'on-off' switching during an alternating magnetic field. The alternating magnetic field triggers the self-generated heat from the incorporated nanoparticles and induces the deswelling of polymer networks in the nanofiber with consequently drug release. The double effects of drug and heat resulted successful for treating human melanoma cells with about the 70% of cancer cell mortality and demonstrated its high potential as manipulative hyperthermia material and switchable drug release platform.[160]

Since cancer cells are known to secrete acids which reduce the pH to below 6.8, Zhao et al. reported in their work the production of a smart drug release system upon stimulation via pH changes. The system was realized by developing polymer fibers embedding drug loaded-mesoporous silica nanoparticles functionalized with $CaCO_3$ as inorganic cap to control the opening of pore entrances of mesoporous silica nanoparticles inside the fibers. $CaCO_3$ starts to dissolve in $Ca^{2+}$ and $CO_2$ gas just after a reduction of the physiological pH in response to an acidic environment. During the release assay, the system demonstrated an improvement of the drug release amount as pH decreased. Moreover, the *in vitro* tests with HeLa confirmed its cytotoxic ability due to the release of drug in acidic environment.[165]

*4.2. Electrospun fibers for gene delivery*

RNA interference (RNAi) is a promising and powerful approach for cancer treatment. It is a mechanism by which double stranded RNAs mediate sequence-specific gene silencing, provided a new tool in the fight against cancer. On the other hand, uncontrolled transgene expression in non-target organs/sites/cells is problematic due to high biological activities of transgene products. Furthermore, undesirable biodistribution of vectors leads to their loss and vector-dependent side effects. Thus, gene delivery systems that are targeted to specific organs/sites/cells are important for not only efficacy but also safety. Since the initial study of Fang and Renneker, demonstrating the possibility of ES pure DNA nanofibers, the use of this technique for a variety of applications has truly exploded.[167]

Electrospun nanofibers mediated siRNA delivery may serve as a potential alternative since the nanofiber can provide sustained long-term delivery at the tumour site.

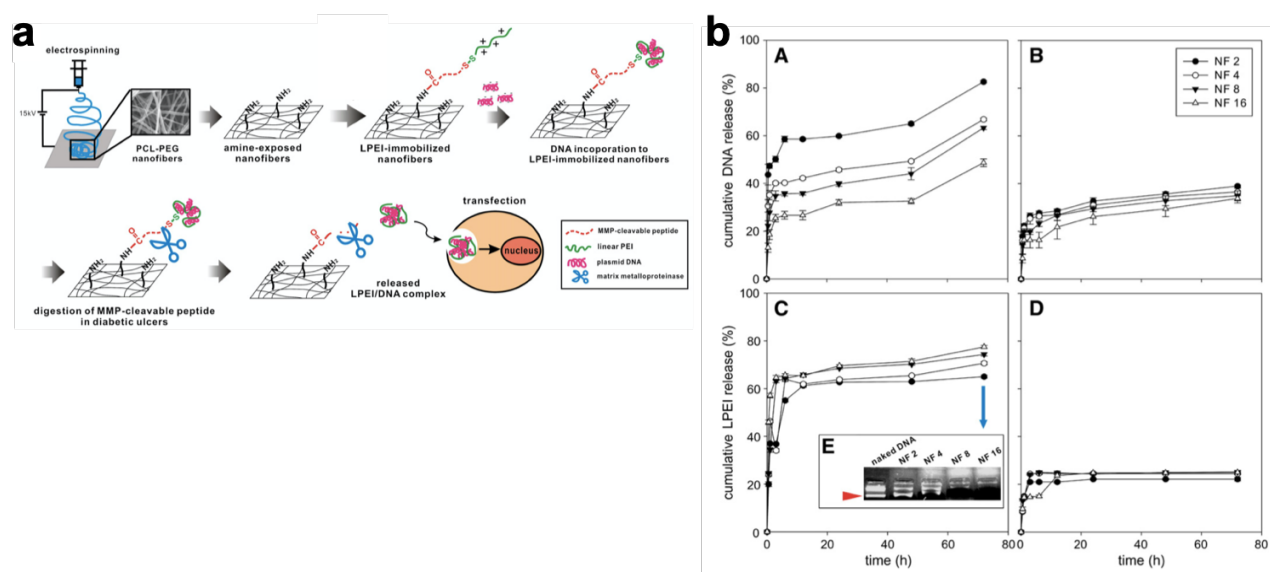

*Figure 6 – (a) Schematic diagram of MMP-responsive electrospun nanofibrous matrix for gene delivery. (b) Release profiles of DNA (A, B) and LPEI (C, D) from nanofibrous matrix in the presence of MMP-2.[168] Reproduced from ref. 168 with permission from [Elsevier], copyright [2010].*

In this regard, Achille and co-workers demonstrated the feasibility of loading into biodegradable fibers a bioactive plasmid encoding for short hairpin (sh) RNA against the cell cycle specific protein, Cdk2. The system leaded to the release of plasmid DNA during the fiber degradation, for over 21 days which suppressed the proliferation of MCF-7 breast cancer cells and affected their viability.[169] Lei and co-workers incorporated a RNAi plasmid designed to specifically suppress MMP-2, an essential proteinase regulating brain tumours invasion and angiogenesis in tumour cells and a cytotoxic drug Paclitaxel into PLGA based fibers to achieve a sustained release of both agents. It was found that DNA nanoparticles complexed with the gene carrier polyethyleneimine and embedded in microfibers were able to inhibit MMP-2 mRNA and protein expression and, particularly, through *in vivo* test on intracranial xenograft tumour model in BALB/c nude mice it was proved that the gene/drug dual delivery microfibers were able to impose significant tumour regression compared with single drug delivery microfibers and commercial drug treatment.[170] We have listed some of the ES nanofiber materials used for gene delivery in cancer cell lines (Table 3).

Table 3 – List of polymer materials used for gene delivery by using ES nanofibers and respective cell lines used for gene delivery analysis.

| Material | Gene particle | Cell line | Ref. |
|---|---|---|---|
| PCL | pCMVb-GFP | 3T3 (Murine fibroblasts cells) | 171 |
|  | miRNA-145 | HuH-7 (Hepatocyte derived cellular carcinoma cells) | 172 |
| PCL-PEG | pEGFP-N1 | NIH3T3 cells | 168 |
| PLLA | pGL3 | COS-7 cells | 14 |
| PEI–PEG | pMSCV | HEK293 cells | 173 |

*Abbreviations: PCL - poly(ε-caprolactone), PEG - poly(ethylene glycol), PLLA- poly(L-lactic acid), PEI – poly(ethylenimine).*

### 4.3. Electrospun fibers for immunosensors/cancer screening

Evidences suggest the detection and the treatment of early stage cancer are essential in improving survival. However, early detection is difficult as it is usually asymptomatic until the disease has spread.[174] Results of previously studies highlighted that circulating antibodies to cancer associated antigens will likely be present in early stages of disease and their detection might be the best way to identify early stage malignancies.[175]

Recent advances in nanoscience and nanotechnology have opened up new horizons for the development of biosensors of enhanced sensitivity, specificity, detection time, and low cost. Sensors miniaturization provides great versatility for incorporation into multiplexed, portable, wearable, and even implantable medical devices. The attractiveness of such nanomaterials relies not only on their ability to act as efficient and stabilizing platforms for the biosensing elements, but also on their small size, large surface area, high reactivity, controlled morphology and structure, biocompatibility, and in some cases electrocatalytic properties. Structuring the transducer at the nanoscale contributes to enlarge the overall surface available for bioreceptor immobilization. Incorporation of nanomaterials into the sensing layers is also often associated with higher mass transfer rates, acceleration and magnification of the transduction process, contributing to signal amplification and faster biosensor response. Especially fibers produced by ES are very interesting in this area since they can be chemically and physically decorated with bioactive molecules like cell-recognizable ligands and provide large surface area.[176] Sensing elements are principally enzymes, antibodies, more scarcely DNA strands or aptamers and the mechanism of action is the electrochemical transduction.

Ali et al. reported the fabrication of an efficient, label-free, selective and highly reproducible immunosensor with unprecedented sensitivity (femto-molar) to detect a breast cancer biomarker for early diagnostics. Mesoporous zinc oxide nanofibers are synthesized by ES technique and fragments of them are electrophoretically deposited on an indium tin oxide glass substrate and conjugated via covalent or electrostatic interactions with a biomarker, the epidermal growth factor receptor 2. Label-free detection of the breast cancer biomarker by this point-of-care device is achieved by an electrochemical impedance technique that has high sensitivity (7.76 kΩ $\mu M^{-1}$) and can detect 1 fM ($4.34 \times 10^{-5}$ ng $mL^{-1}$) concentration. The excellent impedimetric response of this immunosensor provides a fast detection (128 s) in a wide detection test range (1.0 fM–0.5 μM).[177]

Paul et al. reported a novel biosensor platform based on multiwalled carbon nanotubes embedded zinc oxide nanowire for the ultrasensitive detection of carcinoma antigen-125. The system was synthetized by simple ES. The detection label-free was performed by differential voltammetry technique and demonstrated excellent sensitivity (90.14 μA/(U/mL)/$cm^2$) with a detection limit of 0.00113 $UmL^{-1}$ concentration.[178] Soares et al. reported the fabrication of immunosensors based on nanostructured mats of electrospun nanofibers of polyamide 6 and poly(allylamine hydrochloride) coated either with multiwalled carbon nanotubes or gold nanoparticles, whose three-dimensional structure was suitable for the immobilization of anti-CA19-9 antibodies to detect the pancreatic cancer biomarker CA19-9. By using the impedance spectroscopy, the sensing platform was able to detect CA19-9 with a sensitivity of detection limit of 1.84 and 1.57 U $mL^{-1}$ for the nanostructured architectures containing multiwalled carbon nanotubes and gold nanoparticles, respectively. The high sensitivity and selectivity of the immunosensors were also explored in tests with blood serum from patients with distinct concentrations of CA19-9, for which the impedance spectra data were processed with a multidimensional projection technique. The robustness of the immunosensors in dealing with patient samples without suffering interference from analytes present in biological fluids is promising for a simple, effective diagnosis of pancreatic cancer at early stages.[179] More recently, a novel class of pH-sensing ES fibers scaffold was reported by embedding ratiometric pH sensing capsules directly within the lumen of electrospun organic fibers.[180] Upon proton-induced switching, the hybrid ratiometric organic fibers undergo optical changes that could be recorded by fluorescence detectors and correlated to the local proton concentration with micrometer-scale spatial resolution. Biocompatible ES fibers mats with pH sensing properties could be used for *in vivo* spatio-temporal measurements of extracellular acidity of live cells finding potential applications in drug screening and drug discovery for evaluating the efficacy of glycolysis inhibiting anti-cancer drugs in real time.

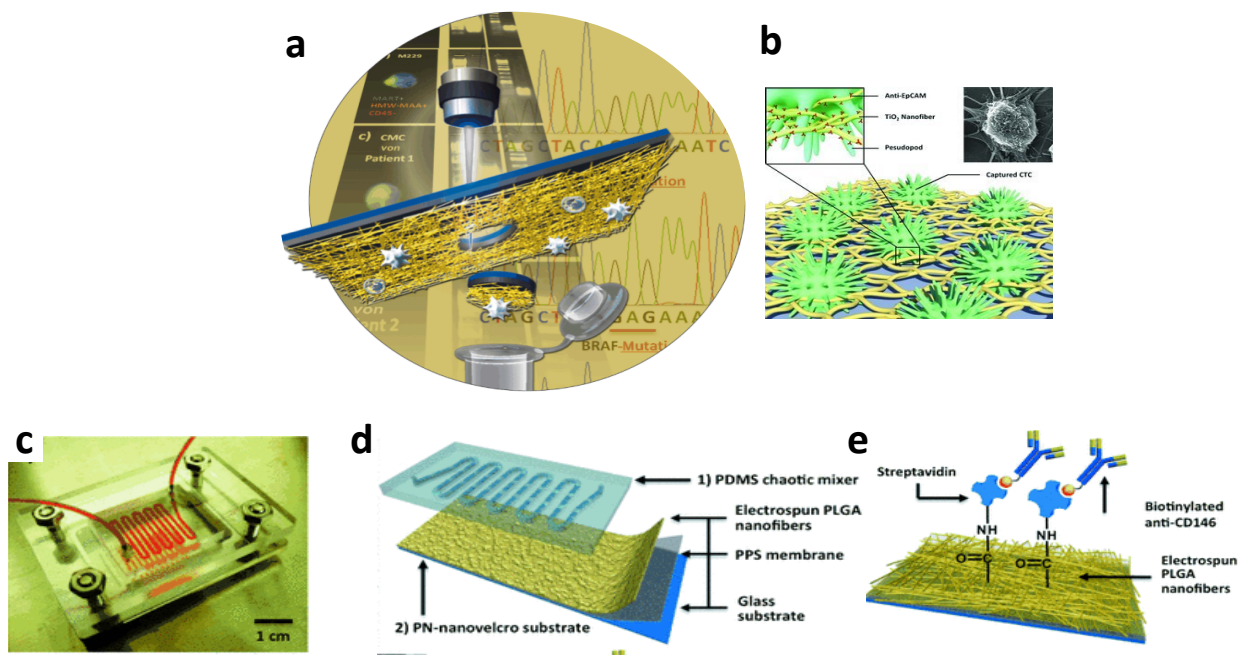

*Figure 7 – (a) A method to detect and isolate single circulating melanoma cells from normal white blood cells with a laser microdissection technique.*[181] *(b) A nanofiber platform TiO$_2$-based and biological cell-capture agent to capture circulating tumour cells.*[182] *(c) Picture of a nanostructured chip for melanoma cell isolation composed by (d) an overlaid PDMS chaotic mixer and a transparent nanovelcro substrate; (e) NHS chemistry is used to covalently anchor streptavidin for conjugation of biotinylated anti-CD146, a melanoma-specific antibody.*[181] *Reproduced from ref. 181 with permission from [John Wiley and Sons], copyright [2013] and ref. 182 with permission from [John Wiley and Sons], copyright [2012].*

## 5. Conclusions

The high versatility, cost effectiveness and scalability of ES is inspiring the design of novel strategies in cancer research. The use of electrospun scaffolds, such as polymer nanofibers that recapitulates the fibrous architecture of native ECM environment, could help to manufacture xeno-free and highly controlled 3D *in vitro* tumour models for cell-based compounds screening, mechanistic studies and pre-clinical drug discovery. Various approaches propose multifunctional and stimuli-responsive electrospun patches for topical release of therapeutics in a controlled and sustained way with improved therapeutics efficacy, reduced side effects and prolonged patients' life expectancy. Electrosupn nanostructured mats also represents attractive sensor platforms to be used for fast diagnosis of cancers at early stages. Although numerous improvements have been achieved by employing electrospun scaffolds in cancer diagnosis and therapy, in depth *in vivo* studies are indispensable before use in clinical trials and medical device products.

**Acknowledgments**

The authors gratefully acknowledge the ERC Starting Grant INTERCELLMED (project number 759959), the My First AIRC Grant (MFAG-2019, No. 22902), My First AIRC Grant (MFAG-2019, No. 22902) and the project "TECNOMED-Tecnopolo di Nanotecnologia e Fotonica per la Medicina di Precisione" [Ministry of University and Scientific Research (MIUR) Decreto Direttoriale n. 3449 del 4/12/2017, CUP B83B17000010001].